\documentstyle[12pt]{article}
\begin{document}
\begin{titlepage}
\begin{flushright}
IASSNS-HEP-95/40
\end{flushright}
\vspace{2.5cm}
\begin{centering}
{\LARGE{\bf Enhancement of the magnetic}}\\
{\LARGE{\bf moment of the electron due}}\\
\vspace{0.2cm}
{\LARGE{\bf to a topological defect}}\\
\bigskip\bigskip
\renewcommand{\thefootnote}{\fnsymbol{footnote}}
Fernando Moraes\footnote{On leave from:\\
Departamento de F\'{\i}sica\\
Universidade Federal de Pernambuco\\
50670-901 Recife, PE, Brazil}\\
{\em School of Natural Sciences\\
Institute for Advanced Study\\
Princeton, NJ 08540\\
U.\ S.\ A.}

\end{centering}
\vspace{1.5cm}

\begin{abstract}

In the framework of the theory of defects/three-dimensional gravitation, it is
obtained a positive correction to the magnetic moment of the electron bound to
a disclination in a dielectric solid.

\end{abstract}

\hspace{0.8 cm}
\end{titlepage}
\def\carre{\vbox{\hrule\hbox{\vrule\kern 3pt
\vbox{\kern 3pt\kern 3pt}\kern 3pt\vrule}\hrule}}

\baselineskip = 18pt

The force between two neutral metallic parallel plates, the Casimir
effect~\cite{Plu}, is a well-known manifestation of the polarization of the
quantum electrodynamical vacuum. The plates change the topology of the space,
thereby modifying the boundary conditions to which the quantized field is
submitted, giving rise to a finite energy density in the space between them.

Another textbook example of quantum electrodynamics is provided by the magnetic
moment of the electron in empty space:
\begin{equation}
\mu=\frac{e\hbar}{2mc}(1+\frac{e^{2}}{2\pi\hbar c}+...).
\end{equation}
Here, the correction to the Dirac value $\mu=\frac{e\hbar}{2mc}$ is due to the
coupling of the electron to the quantized electromagnetic field. Since the
quantum field, even in its vacuum state, is sensitive to the boundary
conditions (or geometry/topology of space) one might ask what happens to the
magnetic moment of an electron between  metallic parallel plates, for example.
The answer~\cite{Bar} is a substantial change in the correction to the Dirac
value of $\mu$. The correction now depends in the relevant parameters related
to the boundary conditions.

A topological defect, like a disclination for instance, changes the boundary
conditions of the quantum field giving rise to a Casimir effect~\cite{Mor} just
like the parallel plate case. Consequently, one should expect a anomalous
magnetic moment for an electron in the neighborhood of such a defect. In this
short note, using a result previously found for cosmic strings~\cite{Mak}, it
is found an explicit expression for the correction $\delta\mu/\mu$ to the
magnetic moment of an electron bound to a disclination in a dielectric medium.

Katanaev and Volovich~\cite{Kat} set up a powerful framework for the study of
elastic media with defects based in three-dimensional gravity theory with
torsion. In fact, they showed the equivalence between the elastic theory of
defects in solids and three-dimensional gravity, allowing for a solid state
interpretation of 3D gravity phenomena. The elastic medium is viewed as a
Riemann-Cartan manifold endowed with a metric that contains the information on
the defects. For example, an elastic medium with a disclination at the origin
is described by
\begin{equation}
ds^2=dz^2+d\rho^2+\rho^2\alpha^2d\varphi^2,
\end{equation}
where $\alpha$ is related to the wedge of dihedral angle $\lambda$, inserted
(or extracted) in the medium in the process of creating the disclination by,
$\alpha=1+\lambda/2\pi$. This metric describes a manifold with singular
curvature at the origin, but otherwise flat. Indeed, the curvature tensor is
given by~\cite{Sok}
\begin{equation}
R^{12}_{12}=R^{1}_{1}=R^{2}_{2}=2\pi\frac{1-\alpha}{\alpha}\delta_{2}(\rho),
\end{equation}
where $\delta_{2}(\rho)$ is the two-dimensional delta function in flat space.
It is clear that $0\leq\alpha<1$ corresponds to a positive-curvature
disclination and $\alpha>1$ to a negative-curvature disclination. $\alpha=1$,
of course, describes the Euclidean medium (absence of disclinations). In
previous articles we have explored this by investigating electronic properties
of disclinated media~\cite{Fur1,Fur2} and the Casimir effect due to a
disclination~\cite{Mor}.

The metric (2) is just the space section of the cosmic string metric
\begin{equation}
ds^2=-c^2dt^2+dz^2+d\rho^2+\rho^2\alpha^2d\varphi^2.
\end{equation}
Here, $\alpha$ depends on the gravitational constant and on the mass density of
the string. This remarkable coincidence allows one to take advantage of the
large amount of literature on physical phenomena related to cosmic
strings~\cite{Vil} for a better understanding of disclinations in solids.

Maki and Shiraishi~\cite{Mak} calculated the contribution to the anomalous
magnetic moment of the electron near a cosmic string to be
\begin{equation}
\frac{\delta\mu}{\mu}=-\frac{(p^2-1)e^2 \hbar}{48\pi^2m^2\rho^2 c^3},
\end{equation}
where $p=1/\alpha$, $e$ is the electron charge, $m$ its mass, and $\rho$ its
distance to the string. In order to use this expression to study the anomalous
moment of the electron near a disclination, a reasonable estimate of $\rho^2$
is needed. This is found as follows.

In our previous work~\cite{Fur1} we showed that electrons are likely to be
bound to negative-curvature disclinations in dielectric media~\cite{Osi}. The
energy eigenstates being given by
\begin{equation}
E=-\frac{m^*e^4\kappa^2(p)}{32\pi^2\hbar^2\varepsilon^2}\frac{1}{(n+\frac{1}{2}+p|l|)^2}+\frac{\hbar^2k^2}{2m^*},
\end{equation}
where $m^*$ is the electron effective mass, $\varepsilon$ is the dielectric
constant of the medium, $n$, $l$ and $k$ are quantum numbers ($n=0,1,2,...$;
$l=0,\pm 1,\pm 2,...$; and $k$ a real number). $\kappa (p)$ is given by
\begin{equation}
\kappa (p)=2\int_{0}^{\infty}\frac{p\coth (px)-\coth (x)}{\sinh (x)}dx.
\end{equation}
At low temperatures, a bound electron will be at the ground state which
corresponds to $n=l=k=0$. The corresponding eigenfunction is~\cite{Fur1}
\begin{equation}
R_{0}= Ce^{-\beta\rho}\,_{1}F_{1}(0,1,\beta\rho),
\end{equation}
where $C$ is a normalization constant,
\begin{equation}
\beta^2 =-2m^{*}E/\hbar^2,
\end{equation}
and $_{1}F_{1}(0,1,\beta\rho)$ is the confluent hypergeometric function which,
in this case~\cite{Sea}, is exactly 1. The mean value of the squared distance
between  the electron and the disclination, for the ground state, is then
easily found to be
\begin{equation}
<\rho^2>=\frac{\int_{0}^{\infty}e^{-2\beta\rho}\rho^3d\rho}{\int_{0}^{\infty}e^{-2\beta\rho}\rho d\rho}=\frac{3}{2\beta^2}.
\end{equation}

Equations (5),  (6), (9) and (10) combined give
\begin{equation}
\frac{\delta\mu}{\mu}=\frac{(1-p^2)\kappa^2(p)e^6}{288 \pi^4 \hbar^3
c^3\varepsilon^2},
\end{equation}
which is then the correction to the magnetic moment of the electron due to the
disclination. Since $0<p<1$ the correction is positive, implying an enhancement
of the magnetic moment in relation to the free electron value. This confirms an
earlier suggestion~\cite{Ben} that negative-curvature disclinations encourage
larger local moments.
\\
\noindent
{\bf Acknowledgment}\\
\noindent
This work was partially supported by CNPq.


\begin{thebibliography}{99}
\bibitem{Plu}G. Plunien, B. M\"uller and W. Greiner, Phys. Rep. {\bf 134}, 87
(1986).
\bibitem{Bar}G. Barton and N.S.J. Fawcett, Phys. Rep. {\bf 170}, 1 (1988)  and
references therein.
\bibitem{Mor}Fernando Moraes, Casimir effect around disclinations, preprint
IASSNS-HEP-95/28  and cond-mat 9505023.
\bibitem{Mak}Takuya Maki and Kiyoshi Shiraishi, Int. J. Mod. Phys. A {\bf 9},
1787 (1994).
\bibitem{Kat}M.O. Katanaev and I.V. Volovich, Ann. Phys. (N.Y.) {\bf 216}, 1
(1992).
\bibitem{Sok}D.D. Sokolov and A.A. Starobinskii, Sov. Phys. Dokl. {\bf 22}, 312
(1977).
\bibitem{Fur1}Claudio Furtado and Fernando Moraes, Phys. Lett. A {\bf 188}, 394
(1994).
\bibitem{Fur2}C. Furtado, B.G.C. da Cunha, F. Moraes, E.R. Bezerra de Mello and
V.B. Bezerra, Phys. Lett. A {\bf 195}, 90 (1994).
\bibitem{Vil}For a recent review see: A. Vilenkin and E.P.S. Shellard, Cosmic
strings and other topological defects (Cambridge Univ. Press, Cambridge 1994).
\bibitem{Osi}Binding of electrons to negative disclinations has been recently
verified numerically, in a gauge theory approach, by V.A. Osipov and S.E.
Krasavin, J. Phys. C {\bf 7}, L95 (1995).
\bibitem{Sea}James B. Seaborn, Hypergeometric Functions and their applications
(Springer-Verlag, New York 1991).
\bibitem{Ben}L.H.Bennett and R.E. Watson, Phys. Rev. B {\bf 35}, 845 (1987).



\end{thebibliography}
\end{document}